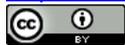

# Toward Efficient Quantum Key Distribution Reconciliation

## Nedra Benletaief, Houria Rezig*, Ammar Bouallegue

Communication System Laboratory Sys'Com, National Engineering School of Tunis, Tunis, Tunisia
Email: benletaief.nedra@gmail.com, houria.rezig@enit.rnu.tn, ammar.bouallegue@enit.rnu.tn

## Abstract

In this paper, we propose how to construct a reconciliation method for the BB84 Quantum Key Distribution (QKD) protocol. Theoretically, it is unconditionally secure because it is based on the quantum laws of physics, rather than the assumed computational complexity of mathematical problems. BB84 protocol performances can be reduced by various errors and information leakages such as limited intrinsic efficiency of the protocol, imperfect devices and eavesdropping. The proposed reconciliation method allowed to weed out these errors by using Turbo codes. Since their high error correction capability implies getting low errors, this method has high performance especially when compared to the last method presented in the literature based on Low-Density Parity Check codes (LDPC). In particular, we demonstrate that our method leads to a significant improvement of the protocol security and of the Bit Error Rate (BER) even with great eavesdropping capability.

## 1. Introduction

Public key cryptosystems based on some computational assumption, especially with large and randomly generated keys, are safe within the context of the current technology. But, their safety can become weaker due to the incredibly high performance of quantum computers and the progress on cryptanalysis.

One solution is to apply QKD which relies on quantum and classical procedures in order to achieve the growing of a secret random string known only to the two parties executing the protocol. The BB84 protocol is one of the widely used QKD protocol that provides a way to share an unconditionally secure key in the presence of an eavesdropper.

This manuscript is organized as follows: In Section 2, we introduce the BB84 protocol. In section 3, we present an overview of previous related works. Especially, we focus on a recently presented method based on the well known LDPC codes. In Section 4, we introduce the basic knowledge of Turbo codes and we describe our proposed reconciliation method. In Section 5, we simulate the efficiency and the correcting ability of our method and give some results according to a pre-established method. Some conclusions and perspectives are consequently drawn in the sixth section.

## 2. Problem Statement: BB84 Protocol

QKD achieves a secret random string namely the key $K = \{0,1\}^E$ known only to the two parties who are executing the protocol. The first protocol for quantum cryptography was proposed in 1984 by Bennett of IBM and Brassard of the University of Montreal and it is called the BB84 protocol [1]. The BB84 protocol like any other protocol of quantum cryptography is based on two principal phases: a quantum phase via a one-way physical quantum channel and a public phase using an authenticated two-way classic ideal channel. A QKD protocol can usually be divided into five steps that will be illustrated bellow (see **Figure 1**).

1) Quantum transmission and reception (see **Table 2**): In order to well explain this step, we use the language of spin $\frac{1}{2}$ as the first experimental demonstration of the protocol in 1991 used the polarization states (see **Table 1**) of single photons to transmit a random key. The protocol uses four quantum states that constitute two bases. A basis is chosen to distinguish the state values without ambiguity.

First, Alice chooses a random string $k_A \in \{0,1\}^F$ of $F$ bits and a random string $QB_A \in \{\oplus, \otimes\}^F$ of $F$ bases, where $F > E$. Then, Alice prepares photons $QS_A$ as a quantum state and sends them to Bob over the quantum channel. Next, Bob measures these incoming states in one of the two bases, chosen at random by using an independent random-number generator from that of Alice.

2) Public comparison: On receiving the state $QS_A$, Bob informs Alice via public channels of the basis $QB_B$ used to accept the states. After that, Alice informs Bob, which states where applied the correct basis. So, they discard the bits corresponding to incorrect choice of bases and use the remaining. This process gives the two parties correlated random variables, called sifted keys $X_A$ and $X_B$ (see **Table 2**).

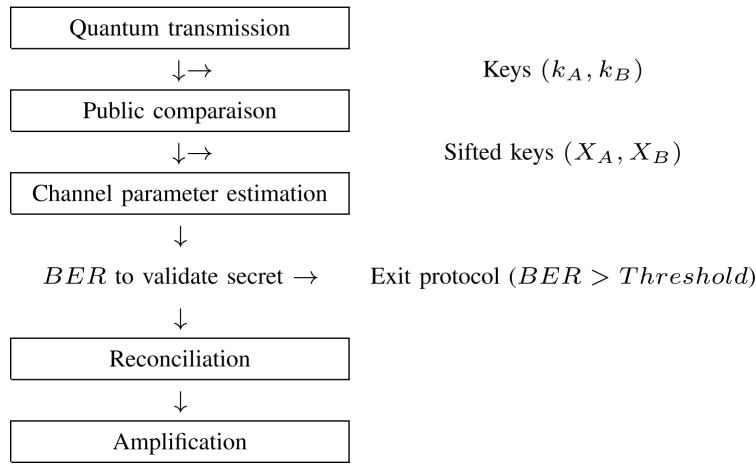

**Figure 1.** Steps of the BB84 protocol.

**Table 1.** Quantum states representation.

| $\oplus$ | $\otimes$ |
|:---:|:---:|
| $0°(\leftrightarrow)$ | $45°(\nearrow)$ |
| $90°(\updownarrow)$ | $135°(\searrow)$ |

3) Channel parameter estimation: In this step, Alice and Bob estimate the channel parameters between them from announced data, in particular the joint probability distribution among Alice, Bob and Eve.

4) Information reconciliation: This step aims to improve BB84 performances by conversation over the public channel. Although the quantum laws of physics provide the BB84 protocol with an unconditionally security proof given by Mayers [2] and later shown by Shor and Preskill [3], the BB84 protocol performance can be reduced by imperfect devices and eavesdropping.

5) Privacy amplification: After error correction, Eve may have partial information of the key. Thus, Alice and Bob need to lower down Eve's information to an arbitrary low value using some privacy amplification protocols. The resulting bits are almost statistically independent of all the information possessed by Eve.

In our work, we are interested in the step of information reconciliation. The construction of an efficient reconciliation scheme and its validation by testing its performance are a difficult problem. Recently, Turbo codes have shown to be good candidates for reconciliation [4]. Before announcing our proposed method based on Turbo codes, we will focus on the most widely investigated related works.

## 3. Related Works

### 3.1. Works on Information Reconciliation

Several methods of information reconciliation for quantum cryptography have already been reported in the literature. As depicted in **Table 3**, we can categorize them into two groups: interactive reconciliation methods and reconciliation methods based on error correcting codes. In what follows, we will only concentrate on the most widely used ones: the Binary algorithm [5], the Cascade algorithm [6], the Winnow algorithm [7], and the LDPC codes based method [8].

**Table 2.** BB84 protocol: example of quantum transmission and public comparison.

| | Alice | Bob |
|---|---|---|
| $k_A$ | 10010001 | |
| $QB_A$ | ⊕⊗⊗⊕⊕⊗⊕⊕ | |
| $QS_A$ | ↕↗↔↘↔↔↗↕ | |
| $QB_B$ | | ⊕⊕⊕⊗⊗⊗⊗⊗ |
| $QS_B$ | | ↕↔↔↘↔↗↕↗ |
| $k_B$ | | 10010001 |
| Base comparison | OK - OK OK OK - - - | |
| Sifted key | $X_A = 1010$ | $X_B = 1010$ |

**Table 3.** Related works.

| Reconciliation | method | Advantages | Disadvantages |
|---|---|---|---|
| Interactive | Binary [5] | -Easy and simple | -Large |
| | Cascade [6] | -Easy and simple | communication |
| | | -Strong ability | overhead |
| | | of error correction | |
| | Winnow | -Communication | -Additional |
| Errors | [7] | time depending | errors (Hamming) |
| correcting | | on the rate | -Great Efficiency |
| codes | LDPC | -Correction of errors | |
| based | [8] | as Cascade | |
| | | -Improvement of the | |
| | | safety of the protocol | |

As interactive reconciliation method, we have binary and cascade. In 1992, Bennett *et al.* proposed a simple and easy reconciliation method called Binary. In 1993, Brassard *et al.* proposed a stronger ability of error correction method called Cascade. Theoretically, we could minimize the information leakage using an interactive protocol. In practice, this would lead to a prohibitively large communication overhead and would limit the effective key rate. That is why, reconciliation methods based on error correcting codes are more practical.

As reconciliation method based on error correcting codes, we have the Winnow algorithm and LDPC codes based method. The Winnow based method was introduced by Buttler *et al.* The winnow algorithm's communication time only depends on the error rate. But, the Winnow introduced additional errors to multiple errors block because the Hamming algorithm can only reveal one single error in each block. Recently, the LDPC codes have shown the ability to correct the same range of errors as Cascade and has the advantage to improve the safety of the used protocol.

In the next subsection, we will describe reconciliation methods based on error correcting codes in more details. A special interest will be attached to the reconciliation method based on LDPC codes.

## 3.2. Reconciliation Methods Based on Error Correcting Codes

### 3.2.1. Overview of the Key Reconciliation Model

In the context of continuous variables, modern coding techniques have been used such as Turbo codes in [9] and LDPC codes in [10] and [11]. In contrast with continuous variable information reconciliation, not much has been done to adapt modern coding techniques to the discrete case. Here, it should be noted that these techniques have the advantage of being very well known and even attaining the Shannon limit for some channels. The main difference between reconciliation and channel coding is that in the case of reconciliation, Alice have no choice of what she sends. She can't restrict her messages to code words of a given code. But, to take advantage of the code formalism (see **Figure 2**), Alice can describe to Bob a code for which her word is a code word. So, a common sequence of bits between Alice and Bob can be shared if Bob guess what code word Alice sent.

To better explain, let $C$ be a linear code and $H$ its parity check matrix. The group $\mathbb{K}_2^E = \{0,1\}^E$ of possible keys sent by Alice can be seen as the product of code words and syndromes: if Alice sends $x$ to Bob, she can tell him the syndrome of $x$, which is $H \cdot x$, thus defining a coset code containing $x$. This coset code is the ensemble $\{y \in \mathbb{K}_2^E \setminus H \cdot y = x\}$.

An equivalent solution is for Alice to randomly choose a code word $U$ from a given code and to send $U \oplus x = \alpha$ to Bob where $\oplus$ represents the addition in the group $\mathbb{K}_2^E$. Bob then computes $y \oplus \alpha$ which allows him to retrieve $U$ if the code is well adapted to the channel between Alice and Bob. Indeed, the side information sent by Alice over the classical authenticated channel corresponds to a change of coordinates allowing one to transform the initial reconciliation problem into the well-known problem of channel coding.

### 3.2.2. Low-Density Parity-Check Codes Based Method

In their work, D. Elkouss *et al.* [12] resort to LDPC codes to accomplish reconciliation. More precisely, the data sample generated by Alice is firstly encoded and then decoded in the Bob's side. LDPC codes will be introduced in the next part of our work.

### 3.3. LDPC Principle

Originally invented by Gallager in the early 1960's [13], LDPC codes have greatly been developed as one of the most promising error correcting codes in the last few years. LDPC codes are a class of error correcting linear block codes that have a sparse parity check matrix, that is with relatively few non zero values limiting decoding complexity. LDPC codes allow communication over noisy channels possible near the Shannon limit. They admit representations in terms of matrix or sparse bipartite graphs, known as Tanner graphs [14], with variable nodes and check nodes. In what follow, we will deal with Tanner graphs as it is the most widely used representation.

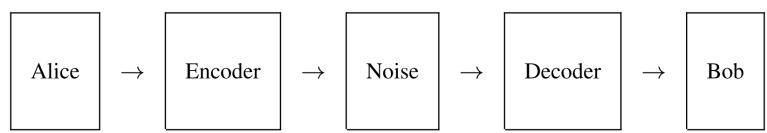

**Figure 2.** Reconciliation method based on error correcting codes.

### 3.4. Coding Scheme of LDPC Codes

A Tanner graph is a useful way to describe LDPC codes that displays the relationship between codeword bits and parity checks.

Parity check matrix (upper) and Tanner graph (lower) representation of a $N = 12$ column, $M = 6$ row, column weight $d_v^{max} = 2$, row weight $d_c^{max} = 4$.

The check nodes represent the set of parity-check equations which define the code. A check node $c_m$ in the graph is called of degree $d_c^m$ if it is connected to $d_c^m$ variable nodes. The variable nodes represent the elements of the codewords. A variable node $v_n$ in the graph is called of degree $d_v^n$ if it is connected to $d_v^n$ check nodes. The graph has an edge between the $n^{th}$ variable node and the $m^{th}$ check node if and only if the $n^{th}$ variable is involved in the $m^{th}$ parity check equation $(H_{mn} = 1)$. Thus, the Tanner graph is a graphical depiction of the parity check matrix (see **Figure 3** for example).

A family of LDPC codes is defined by two generating polynomials $\lambda(x)$ and $\rho(x)$. We denote by $\lambda_i$ the fraction of edges which are connected to variable nodes of degree $d_v^i$ and $\rho_i$ the fraction of edges which are connected to check nodes of degree $d_c^i$.

$$\lambda(x) = \sum_{i=2}^{d_v^{max}} \lambda_i x^{i-1}.$$

$$\rho(x) = \sum_{i=2}^{d_c^{max}} \rho_i x^{i-1}.$$

### 3.5. Decoding Scheme of LDPC Codes

Decoding method is based on belief propagation. Decoding is achieved by the exchange of messages. Each node generates and propagates messages to its neighbors on the other side of the bipartite graph connected by an edge based on its current incoming messages. The exchanged messages correspond to the loglikelihood ratio (LLR) of the probabilities of the bits. The sign of the LLR indicates the most likely value of the bit and the absolute value of the LLR gives the reliability of the message. The generating functions of the two messages (variable-to-check message and check-to-variable message) are defined as:

$$m_{c_a \to v_j} = \log\left( \frac{P\left(E_{c_a} = 0 \middle| v_j = 0, r\right)}{P\left(E_{c_a} = 0 \middle| v_j = 1, r\right)} \right)$$

and

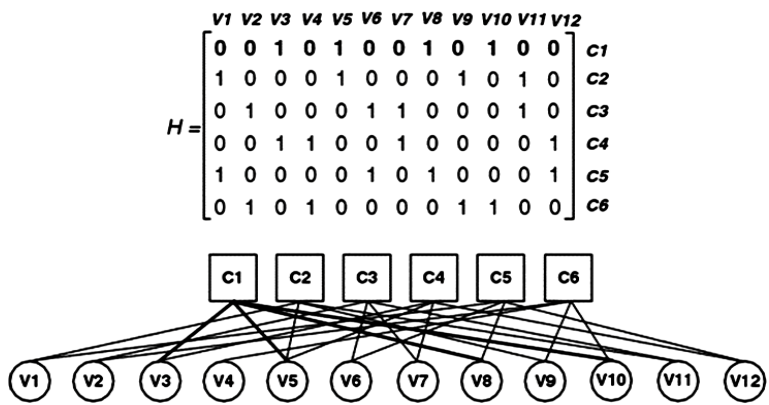

**Figure 3.** Parity check matrix (upper) and Tanner graph (lower) representation of a $N = 12$ column, $M = 6$ row, column weight dmax $v = 2$, row weight dmax $c = 4$.

$$m_{v_j \to c_a} = \log \left( \frac{P\left(v_j = 0 \middle| r, \left\{ E_{c_a} = 0, c_a \in N\left(v_j\right) \setminus c_j \right\} \right)}{P\left(v_j = 1 \middle| r, \left\{ E_{c_a} = 0, c_a \in N\left(v_j\right) \setminus c_j \right\} \right)} \right)$$

where $r$ is the received signal and $E_{c_a} = \sum_{v_n \in N(c_a)} v_n$ is the check equation.

In general, the iterative procedure of belief propagation decoding is composed of two steps.

At the $i^{th}$ iteration, in the first step, all values of the check-to-variable messages are updated by using the values of the variable-to-check messages obtained at the $(i-1)^{th}$ iteration (see **Figure 4(a)**). In the second step, all values of the variable-to-check messages are updated by using the values of the check-to-variable messages newly obtained at the $i^{th}$ iteration (see **Figure 4(b)**). This iterative process continues until the stopping rule is satisfied.

After giving a general presentation of related works, we introduce our method in the next section.

## 4. Our Turbo Codes Based Method

In our work, we propose to resort to Turbo codes to accomplish reconciliation. For this reason, we dedicate this section to explore the Turbo codes theoretical principles.

### 4.1. Turbo Codes Principle

In 1993, an approach to error correction coding was introduced with performance attaining the Shannon limit and providing for very long codewords relatively modest decoding complexity. These codes were termed Turbo codes by their inventors Berrou and *et al.* [15].

#### 4.1.1. Coding Scheme of Turbo Codes

The codes are constructed by using two or more component codes on different interleaved versions of the same information sequence. The constituent codes are usually two identical recursive systematic convolutional codes (see **Figure 5**).

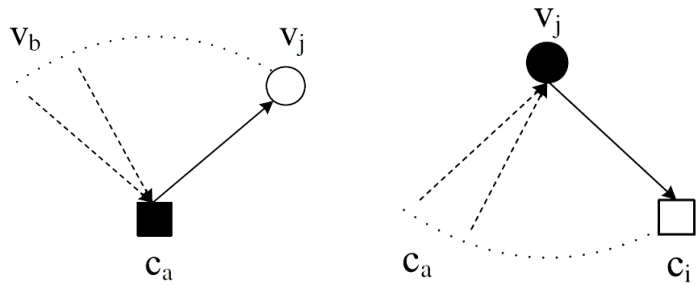

**Figure 4.** Message passing of belief propagation decoding.

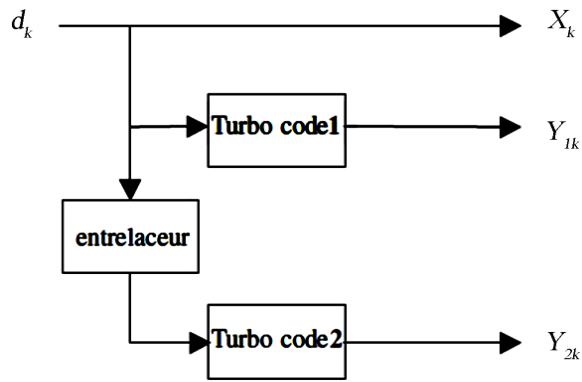

**Figure 5.** The generic Turbo encoder.

The input sequence to be encoded is divided into blocks of same length. Each block is encoded by the first encoder and interleaved before passing through the second encoder.

### 4.1.2. Decoding Scheme of Turbo Codes

The Turbo decoder consists of two, or more, Soft-In Soft-Out (SISO) maximum likelihood decoders (see **Figure 6**).

For conventional codes, the final step at the decoder yields hard-decision decoded bits or decoded symbols. Whereas, for a concatenated scheme as its case of Turbo code, the decoding algorithm don't pass hard decisions among the decoders. But, to best exploit the information learned from each decoder, the decoding algorithm makes an exchange of soft decisions. Decoders are operating in parallel, exchanging iteratively extrinsic information. In other words, the concept behind Turbo decoding is to pass soft decisions from the output of one decoder to the input of the other decoder, and to iterate this process several times so as to produce more reliable decisions. Two families of decoding algorithms are commonly used in Turbo decoding: Soft Output Viterbi Algorithms (SOVA) and Maximum A Posteriori (MAP) algorithms. The MAP algorithm is more efficient than the SOVA. The MAP algorithm aims of computation of the Logarithm of Likelihood Ratio LLR relative to $d_i$:

$$\Lambda(d_i) = \ln\left(\frac{Pr\{d_i = 1 | R_1^i\}}{Pr\{d_i = 0 | R_1^i\}}\right)$$

where $R_1^i$ is the noisy received sequence. The principle of MAP algorithm is to process separately data available between steps 1 to $i$ and between steps $i+1$ to $k$ by introducing forward state probabilities $\alpha_i(m)$ and backward state probabilities $\beta_i(m)$.

$$\alpha_i(m) = Pr\{S_i = m, R_1^i\}.$$

$$\beta_i(m) = Pr\{R_{i+1}^k | S_i = m\}.$$

### 4.1.3. Interleaving

The efficiency of the Turbo codes is due to the use of an iterative process at the decoder side and the presence of an interleaver at the encoder side. The interleaver adds randomness-like effect to the code and give the greater free distance as possible to the concatenated code.

## 5. Results

### 5.1. Simulation Conditions

In this section, we discuss the experimental performances of our reconciliation method. We have implemented a QKD protocol as described in [1]. Also, we have implemented and experimented our reconciliation method on the case of a special type of eavesdropping strategy: Intercept and Resend.

Intercept and Resend is the most known eavesdropping individual strategy that can be implemented with actual technology means.

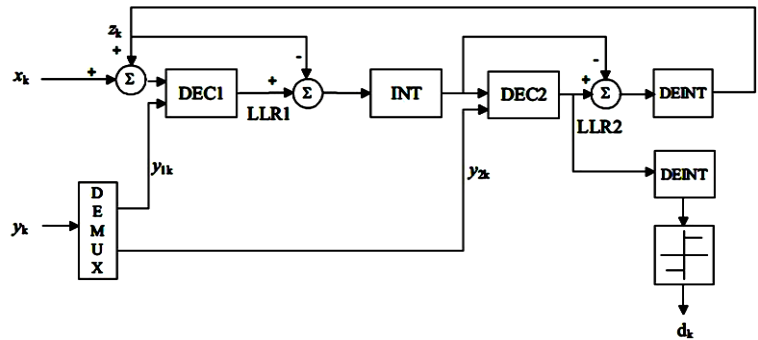

**Figure 6.** The generic Turbo decoder.

As described by **Table 4**, Intercept and Resend attack is where Eve measures the quantum states sent by Alice. Then, she pretends as Alice and sends replacement states to Bob corresponding to her measurement result prepared in a chosen base from the two possible bases.

Intercept and Resend attack on BB84 protocol: example of transmission and reception this produces errors in the key Alice and Bob share. As Eve has no knowledge of the basis a state sent by Alice is encoded in, she can only guess which basis to measure in, in the same way as Bob. If she chooses correctly, she measures the correct photon polarization state as sent by Alice, and resents the correct state to Bob. However, if she chooses incorrectly the base, the state sent to Bob can't be the same as the one sent by Alice. Then, even if Bob measures this state in the Alice's basis, he gets a random result because Eve has sent him a state in the opposite basis.

We define for this eavesdropping strategy the probability $s$ that one qubit sent by Alice is eavesdropped.

## 5.2. The Security of the Proposed Method

In contrast to traditional public key cryptography which relies on the computational difficulty of certain mathematical functions, the security of quantum cryptography relies on the foundations of quantum mechanics. We have now collected all the ingredients to describe the results by studying secure information which is an important parameter defined by [16]:

$$I_s = I_{AB} - I_{AE}$$

where $I_{AB}$ is the amount of information exchanged between Alice and Bob and $I_{AE}$ is the one exchanged between Alice and Eve.

$$I_{AB} = \log_2\left(2 - \frac{s}{2}\right) - \frac{s}{4}\log_2\left(\frac{4}{s} - 1\right).$$

$$I_{AE} = \frac{1}{2}\log_2\left(2 - \frac{s^2}{4}\right) + \frac{s}{4}\log_2\left(\frac{2+s}{2-s}\right).$$

We represent in **Figure 7** secure information $I_s$ as function of *s*-values for intercept and resend eavesdropping.

**Table 4.** Intercept and resend attack on bb84 protocol: example of transmission and reception.

|  | Alice | Eve | Bob |
|---|---|---|---|
| $k_A$ | 10010001 |  |  |
| $QB_A$ | ⊕⊗⊕⊗ |  |  |
|  | ⊕⊕⊗⊕ |  |  |
| $QS_A$ | ↕╱↔╲↔↔╱↕ |  |  |
| $QB_E$ |  | ⊗⊕⊕⊗ |  |
|  |  | ⊕⊕⊕⊗ |  |
| $QS_E$ |  | ╱↔↔╲↔↕↕╱ |  |
| $k_E$ |  | 00010110 |  |
| $QB_B$ |  |  | ⊕⊕⊕⊗ |
|  |  |  | ⊕⊗⊕⊗ |
| $QS_B$ |  |  | ↔↔↔╲↔╱↕╱ |
| $k_B$ |  |  | 00010010 |
| Base | OK - OK OK |  |  |
| comparison | OK - - - |  |  |
| Sifted key | $X_A = 1010$ |  | $X_B = 0010$ |

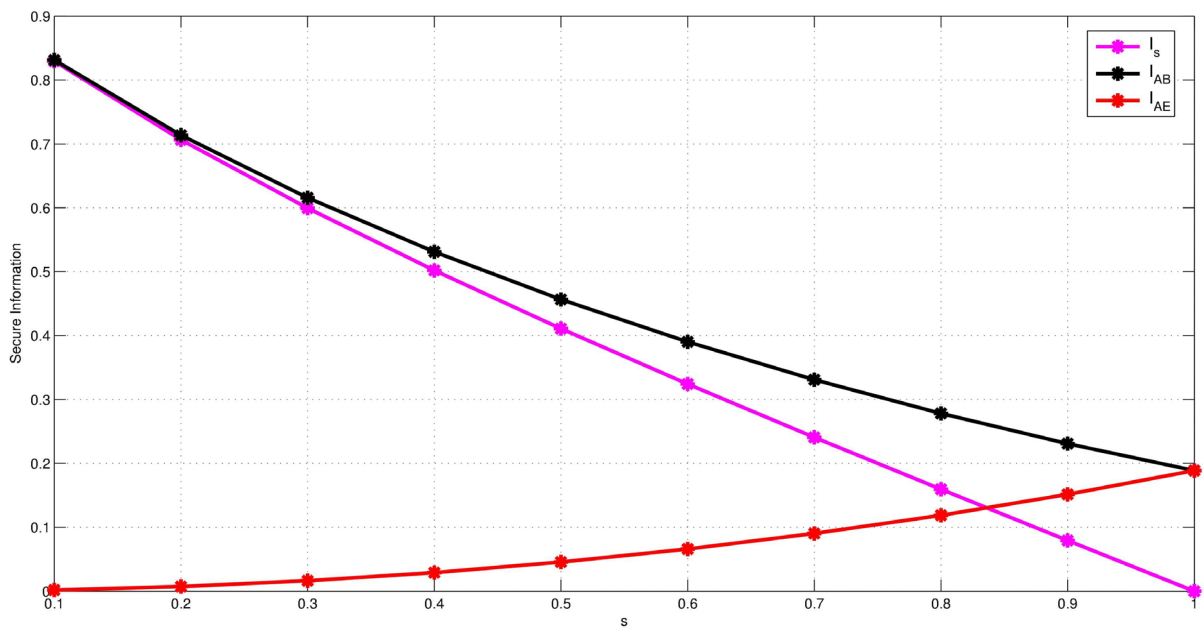

**Figure 7.** Secure Information as function of *s*-values for Intercept and Resend eavesdropping.

It can be noted that secure information decreases up to 0 as the parameter of eavesdropping attack increases to 1. This means that when Eve accedes to all Alice's states, Alice and Bob will no longer have a secret and that is why the protocol should be stopped.

Based on the last equations, we can express the bound of the secure information $I_s$ by:

$$I_s \leq 1 + (1-\delta)\log 2(1-\delta) + \delta \log 2(\delta)$$

where the parameter $\delta$ corresponds to the decoding error probability.

**Figure 8** shows that Turbo codes can improve the security of the BB84 protocol by increasing the secret information better than LDPC codes. In fact, we see for example that the secure information for an eavesdropping probability of 0.5 is equal to 0.4109 without reconciliation. It increases slightly to 0.4179 thanks to the application of the reconciliation method based on LDPC codes and it soar up to 0.6946 after introducing our method.

## 5.3. The Error Correction Ability of the Proposed Method

The performance of the reconciliation method can be evaluated by measuring the Bit Error Rate. In **Figure 9**, the effect of the eavesdropping attack and the performance of our reconciliation method are examined. The BER of the BB84 protocol decoupling with Turbo coding reconciliation scheme is shown for different values of the eavesdropper parameter $s$. As a reference, we also show the BER of the reconciliation method based on LDPC codes and the simulated and theoretical BER without application of the reconciliation method given by [4]:

$$BER = \frac{s}{4}.$$

It can be noted that BER decreases up to 0 as the parameter of eavesdropping attack s decreases also to 0. This means that when Eve has a little access to Alice states, Alice and Bob will have the same sifted keys because we have made the assumption that the channel is perfect. And, as the eavesdropper capability to catch Alice's states increases, we have an increasing BER as shown in **Figure 9**. Looking at the value of the parameter $s$ equal to 0.1, we see that the achieved BER amounts to 0.025 for simulated BB84 protocol and decreases to 0.0018 thanks to the reconciliation by LDPC codes and to 0.0014 thanks to our proposed method, which means that it was devised by 13.89 by the application of the first method and by 17.85 by the introduction of our method of reconciliation. The gain decreases when the eavesdropper capability increases. Indeed, the achieved BER after reconciliation based on LDPC codes is 0.1852 and 0.1223 after application of our method for $s$ equal to 0.9 and 0.225 given by simulation of BB84 protocol without reconciliation, which means that it was devised by 1.21 and 1.83 respectively for the pre-established method and our method (**Table 5**).

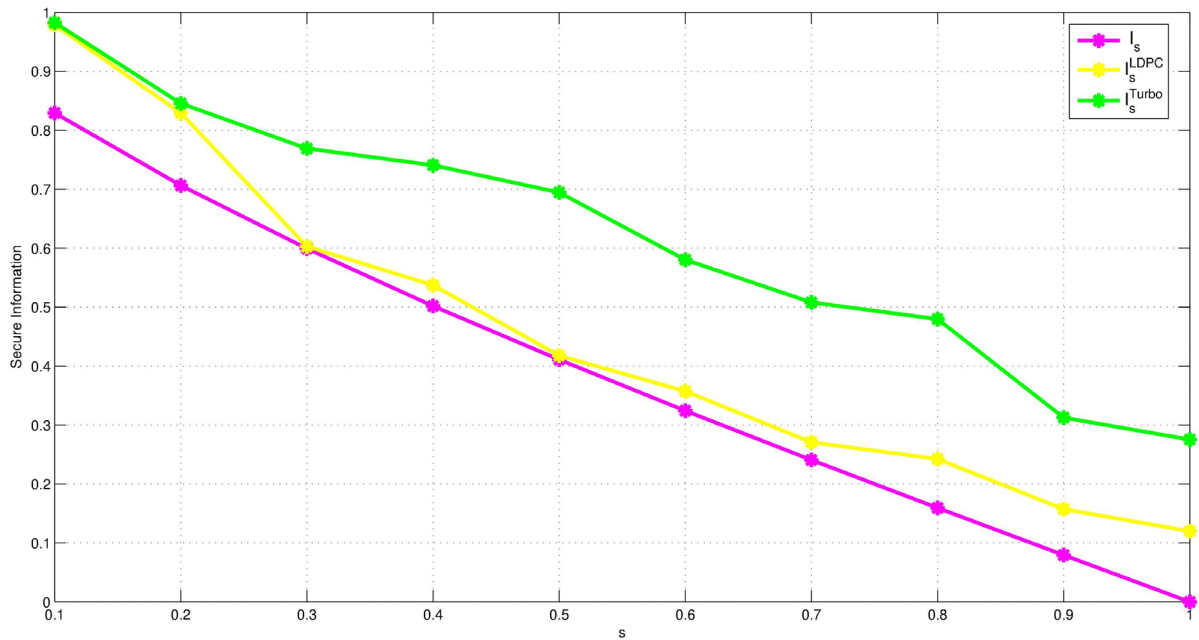

**Figure 8.** Secure Information as function of *s*-values for Intercept and Resend eavesdropping (without reconciliation and with reconciliation with Turbo codes and LDPC codes).

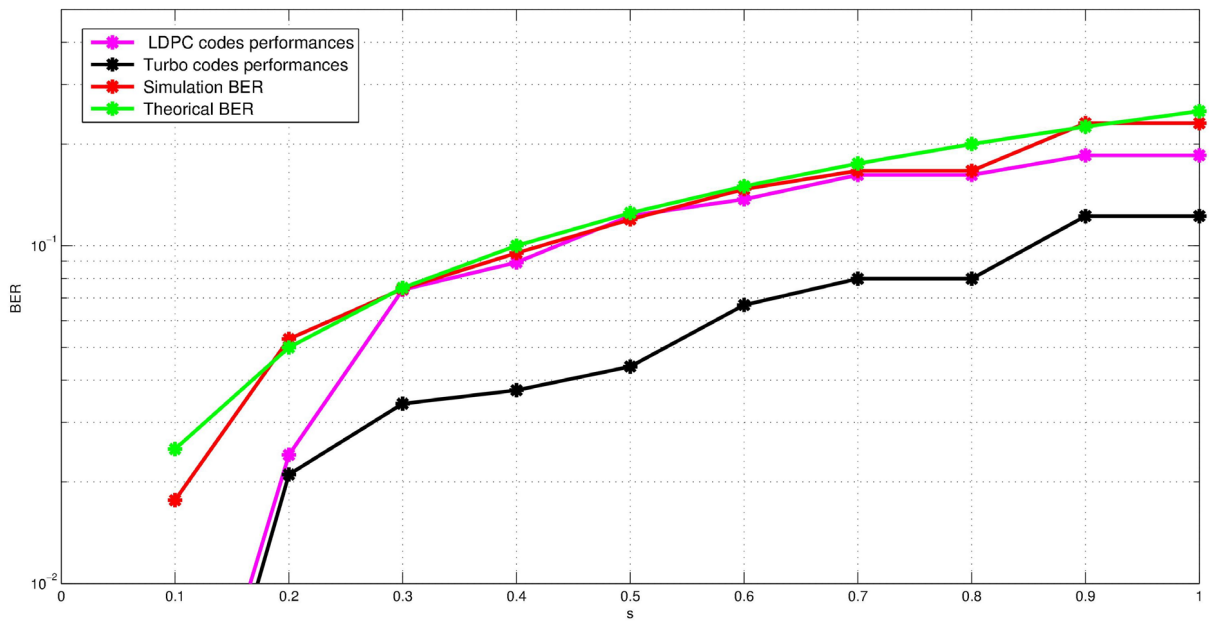

**Figure 9.** Bit Error Rate as function of *s*-values for Intercept and Resend eavesdropping.

**Table 5.** Ber for intercept and resend attack on bb84 protocol.

|  | $BER_{theo}$ | $BER_{LDPC}$ | $BER_{Turbo}$ |
|---|---|---|---|
| $s = 0.1$ | 0.025 | 0.0018 | 0.0014 |
| $s = 0.4$ | 0.1 | 0.0892 | 0.0374 |
| $s = 0.9$ | 0.225 | 0.1852 | 0.1223 |

To sum up, we can mention that we have proven by simulation the efficiency of our reconciliation method as the average of initial Bit Error Rate equal to 0.13 decreases to 0.11 after reconciliation based on LDPC codes and 0.06 after reconciliation based on Turbo codes. Thus, our reconciliation method succeeds in removing on average more than 50% of the errors caused by eavesdropping. One should note that the maximum Bit Error Rate admissible to distribute a secret is 0.11 which is obtained with our method of reconciliation even with a great eavesdropping capability.

Two other criteria that one should keep in mind when evaluating a reconciliation method: its complexity and its rapidity. This last criterion is especially relevant in the case of highly interactive schemes where latency can become an issue. We can say that the integration of our method of reconciliation doesn't affect so much the simulation time of the BB84 protocol. Just to have some order of magnitude, we give the BB84 protocol time of running which is about 0.076 s and the reconciliation method demands 0.023 s. This shows again that reconciliation can be advantageously solved with Turbo codes.

## 6. Conclusion

The BB84 protocol allows remote parties to share secret keys. But, the keys generated by this protocol will contain some errors. Such a situation, that the legitimate partners must remove the errors by public discussion called reconciliation. We proposed an efficient reconciliation method. Our theoretical model assumptions are supported by experimental results, confirming the superiority of our method that outperform the LDPC codes based method which seems the most efficient method when this manuscript is written. There are many related works that are worth further investigation. We can generalize our research by taking into consideration not only eavesdropping effect, but also channel noise and other unwanted interactions in quantum computation and communication.